\documentstyle[twoside,fleqn,espcrc2,epsfig,amssymb]{article}

\newcommand{\vg}{\gamma(P^2)}
\title{Parton Content of Real and Virtual Photons}

\author{I. Schienbein\address{Institut f\"ur Physik, Universit\"at 
        Dortmund\\ 
        D--44221 Dortmund, Germany (e-mail: schien@hal1.physik.uni-dortmund.de)}%
}
       
\begin{document}

\begin{abstract}
Parameter--free and perturbatively stable leading order (LO) and next--to--leading
order (NLO) parton densities for real and virtual photons are presented.
\end{abstract}

\maketitle

\section{Introduction}
The partonic content of real and virtual photons can be measured, for
example, in electron positron scattering.
The measured $e^+ e^-$ cross section can be obtained by a convolution
of a flux of target photons \cite{EPA} with the cross section of deep inelastic
electron photon scattering 
\begin{eqnarray}
d\sigma(e e\rightarrow e X) &=& f_{\vg/e} * d\sigma(e \vg \rightarrow
e X)\nonumber \\
&& + {\cal O}(P^2/Q^2)\ .
\end{eqnarray}
The virtuality of the target photon, $P^2$, has to be much smaller than the
virtuality $Q^2$ of the probe photon
such that the neglected terms of
order ${\cal O}(P^2/Q^2)$ are small.

In the following, we are interested in the photon structure function 
$F_2^{\vg}(x,Q^2)$ which enters the 
deep inelastic electron photon cross section in exactly the same way as
is well known from deep inelastic electron proton scattering
\begin{eqnarray}
\frac{d\sigma(e \vg \rightarrow e X)}{dx dQ^2} &\propto& (1+(1-y)^2) 
F_2^{\vg}\nonumber\\
&& - y^2 F_L^{\vg}\ .
\end{eqnarray}

$F_2^{\vg}$ is a convolution of the {\em massless} parton densities 
of the real or virtual
photon with the respective {\em massless}
Wilson coefficients, in order to obtain a factorization scheme 
independent, \mbox{i.\ e.,} physically meaningful result \cite{GRSvg99}
\begin{eqnarray}
F_2^{\vg}(x,Q^2) &=& \sum_{u,d,s} 2 x e_q^2 \times \Bigg[q^{\vg}\nonumber\\
&+& \frac{\alpha_s}{2 \pi} (C_q * q^{\vg} + C_g * g^{\vg})\nonumber\\
&+& \frac{\alpha}{2 \pi} e_q^2 C_\gamma \Bigg]\ .
\label{F2}
\end{eqnarray}
This also holds for the direct photon coefficient $C_\gamma$, \mbox{i.\ e.,}
$C_\gamma$ has to be the massless direct photon coefficient 
(irrespective of $P^2$)
since it is related to the massless photon to quark and photon to gluon
splitting functions 
$P_{q\gamma}^{(0)}$ and $P_{g\gamma}^{(0)}$ in the evolution equations.

As usual, we work in NLO in the DIS$_\gamma$ scheme in which the 
destabilizing $\log (1-x)$ of $C_\gamma$, as calculated in the 
${\rm \overline{MS}}$ scheme in (\ref{F2}), is absorbed
into the quark distributions
\begin{eqnarray}
(q+\bar{q})_{\rm{DIS}_{\gamma}}^{\gamma} & = & 
   (q+\bar{q})_{\overline{\rm{MS}}}^{\gamma} + \frac{\alpha}{\pi}\,
       e_q^2\, C_{\gamma}^{\overline{\rm{MS}}}(x)\nonumber\\
g_{\rm{DIS}_{\gamma}}^{\gamma} & = & g_{\overline{\rm{MS}}}^{\gamma}\ .
\end{eqnarray}
Again, it should be emphasized that we use the {\em real photon} coefficient
irrespective of $P^2$.

\section{Boundary Conditions}
The parton distributions are then obtained by QCD evolution of 
appropriate boundary conditions at a low scale $Q_0^2\approx 0.3$ GeV$^2$, 
which will be presented in the following.
The exact LO and NLO values of the universal (i.\ e.\ hadron--independent) input
scale $Q_0$ are fixed by the experimentally well constrained radiative 
parton densities of the proton \cite{GRV98}.

\subsection{Real Photon}
\begin{figure}[thb]
\epsfig{figure=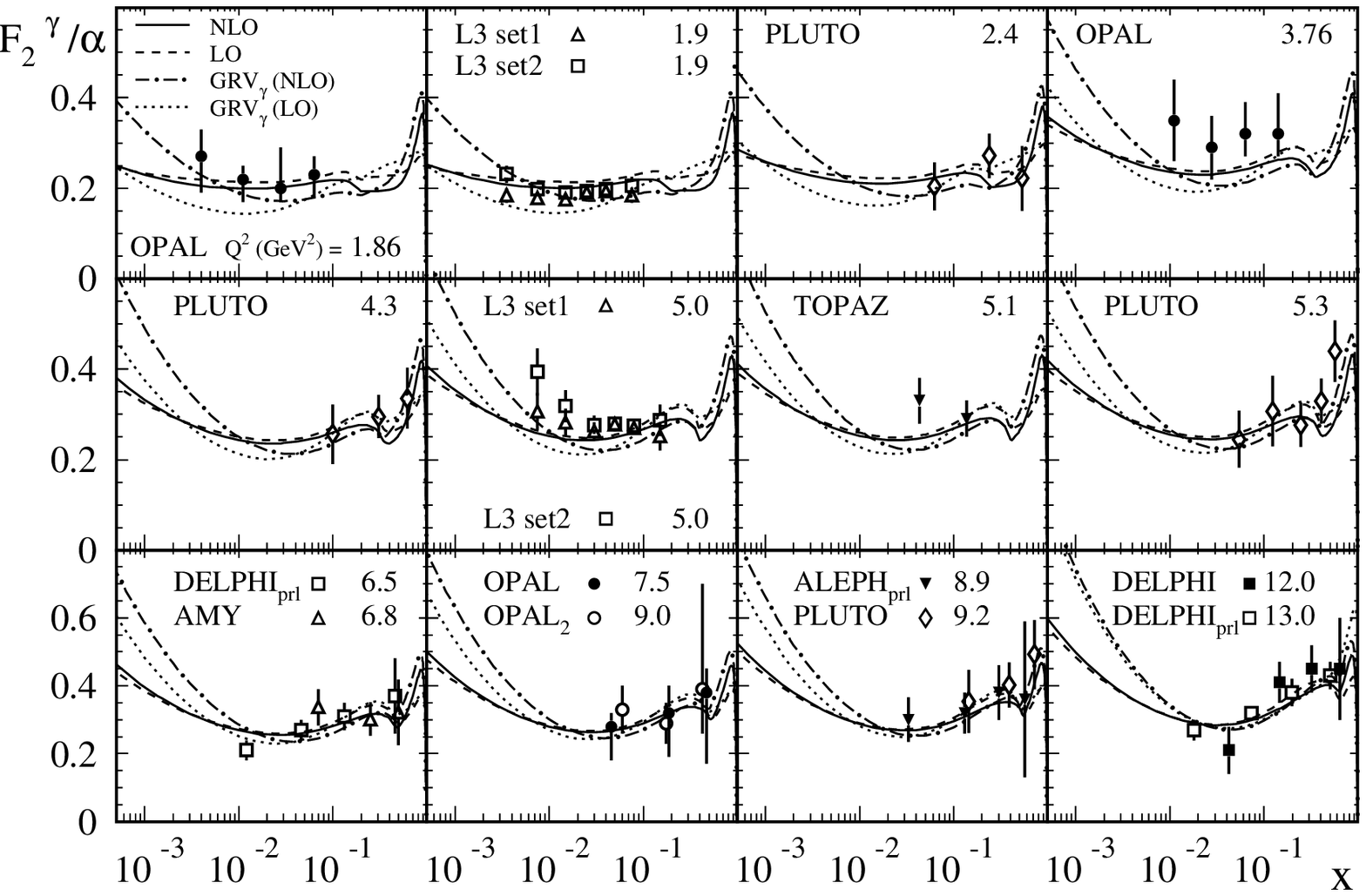,width=7.5cm}
\epsfig{figure=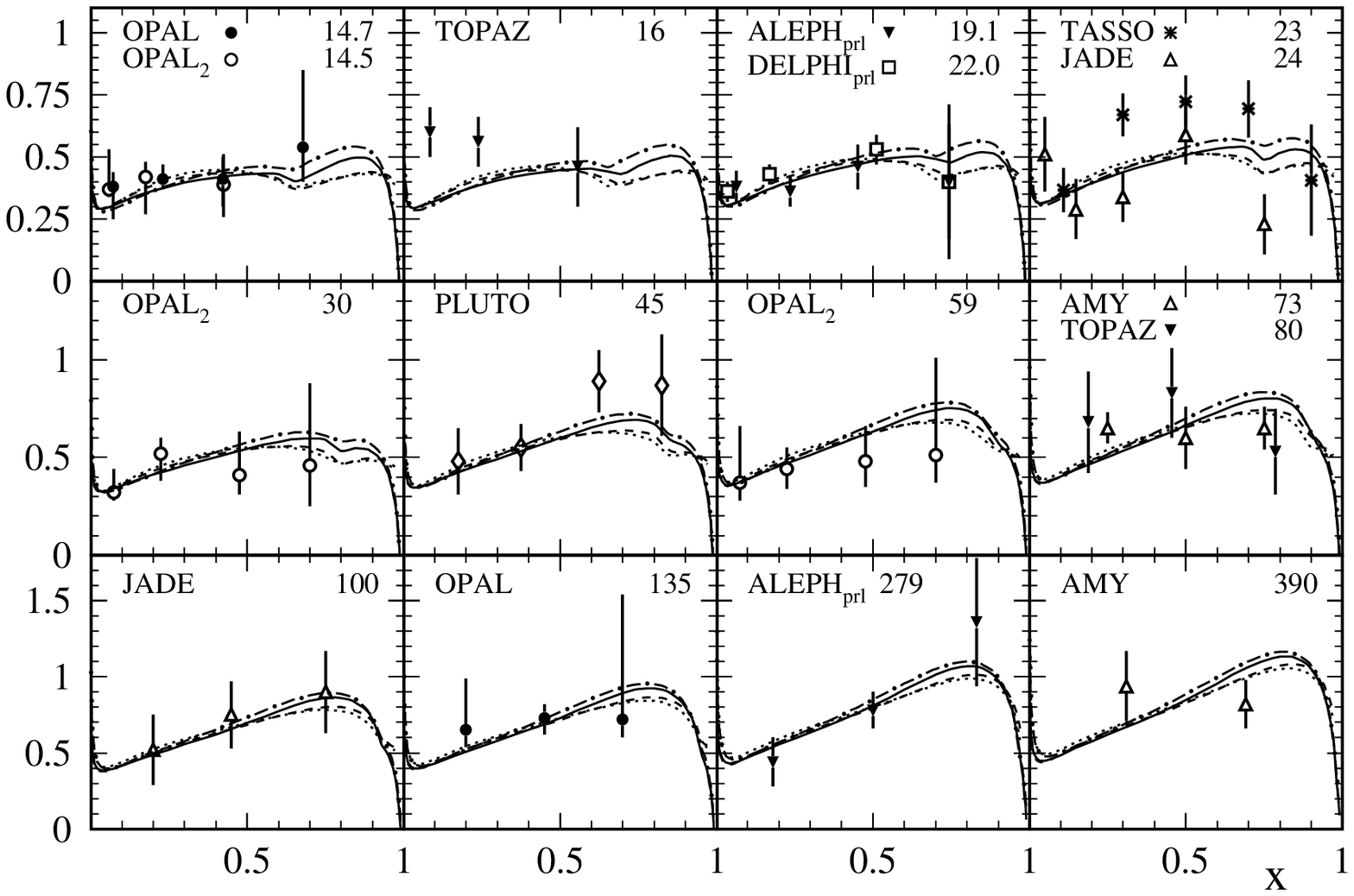,width=7.5cm}

\vspace*{-0.8cm}
\caption{\small
	Comparison of our radiatively generated LO and 
	NLO(DIS$_{\gamma}$) predictions for $F_2^{\gamma}(x,Q^2)$, based 
	on the valence--like parameter--free VMD input in eq.\ (\ref{input}), with
	the data of ref.\ \cite{Data99}.  For comparison the 
	GRV$_{\gamma}$ \cite{GRVg92} results are shown as well.  In both 
	cases, the charm contribution has been added, in the relevant 
	kinematic region $W\geq 2m_c$, according to fixed order perturbation 
        theory.}
\label{fig1b}
\end{figure}

The boundary condition for the real photon \cite{GRSvg99} is given
by a vector meson dominance (VMD) ansatz where (at the low scale $Q_0$)
the physical photon
is assumed to be a coherent superposition of vector mesons which
have the same quantum numbers as the photon
\begin{eqnarray}
f^{\gamma}(x,Q_0^2) = f_{had}^{\gamma}(x,Q_0^2) 
= \alpha G_f^2 f^{\pi^0}(x,Q_0^2)
\label{input}
\end{eqnarray}
with $G_{u,d}^2 = (g_{\rho} \pm g_{\omega})^2$ and $G_g^2 = G_s^2 
= g_{\rho}^2 + g_{\omega}^2$ ($g_{\rho}^2=0.50$, $g_{\omega}^2=0.043$). 
This optimal coherence maximally enhances the up quark which is favoured by the 
experimental data.

Since parton distributions of vector mesons ($\rho, \omega$,...) are unknown
we furthermore assume that these are similar to pionic parton distributions.
Thus, for given pionic parton distributions our model has no free parameter.

\subsection{Pion}
Since only the pionic valence density is experimentally
rather well known, we utilize a constituent quark model to relate
the pionic light sea and gluon to the much better known parton distributions
of the proton \cite{GRV98}.
In this model the proton and pion are described by scale independent 
constituent distributions ($U^p$, $U^\pi$,...)
convoluted with their universal partonic content $f_c$ \cite{Alt74}
\begin{eqnarray}
f^p & = & \int_x^1\frac{dy}{y}\left[ U^p(y)+D^p(y)\right] f_c
\left( \frac{x}{y}, Q^2\right)\\
f^{\pi} & = & \int_x^1\frac{dy}{y} \left[ U^{\pi^+}(y)+
  \bar{D}\,^{\pi^+}(y)\right] f_c \left( \frac{x}{y}, Q^2\right)
\nonumber
\end{eqnarray}
where $f=v,\,\bar{q},\,g$.

Since in Mellin-$n$-space the convolution is a simple product, the ratio  
$f^{\pi}/f^p$ 
is independent of the flavour $f$ and one can easily find
boundary conditions for the pionic gluon and sea  which only depend on the rather
well determined valence distribution of the pion and the parton distributions 
of the proton \cite{GRS98,GRSpi99}
\begin{equation}
g^{\pi}(n,Q_0^2) = \frac{v^{\pi}}{v^p}\, g^p,
\quad\quad
\bar{q}\,^{\pi}(n,Q_0^2) = \frac{v^{\pi}}{v^p}\,
\bar{q}\,^p.
\end{equation}

\subsection{Virtual Photon}
In this subsection boundary conditions for the much more
speculative virtual photon are proposed.

First recall that
the photonic parton distributions can be written as a sum of a 
pointlike and a hadronic part
\begin{equation}
f^{\vg}(x,Q^2) = f_{pl}^{\vg}(x,Q^2) + f_{\rm{had}}^{\vg} 
  (x,Q^2)\, .
\label{plhad}
\end{equation}
The pointlike part is a particular solution of the inhomogenous evolution
equations of the photon and vanishes by definition at the input scale 
$\tilde{P}^2 \equiv \max(P^2,Q_0^2)$:
$f^{\gamma (P^2)}_{pl}(x,\tilde{P}^2) = 0$.
While the pointlike solution is perturbatively calculable, 
the homogenous (hadronic) solution requires a boundary condition.
At $\tilde{P}^2$ we assume the hadronic part of the virtual photon
to be given by the hadronic part of the real photon 
suppressed by a rho-meson propagator
$\eta(P^2)=(1+P^2/m_\rho^2)^{-2}$:
\begin{equation}
f_{\rm{had}}^{\gamma(P^2)}(x,\tilde{P}^2) = \eta(P^2) 
f_{\rm{had}}^{\gamma}(x,\tilde{P}^2)
\end{equation}
This boundary condition of course smoothly extrapolates to the real photon
case.

Note that, the employed dipole suppression factor is somewhat speculative and can
be regarded as the simplest choice of modelling the $P^2$--suppression.

\begin{figure}[htb]
\vspace*{-0.25cm}
\epsfig{figure=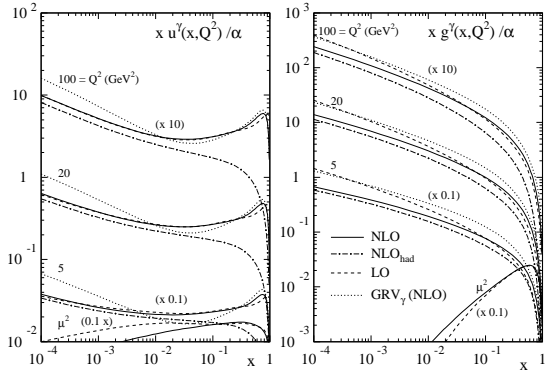,angle=0,width=7.5cm}

\vspace*{-0.8cm}
\caption{\small
        Detailed small--$x$ (as well as large--$x$) 
	behavior and predictions of our radiatively generated 
	$u^{\gamma}=\bar{u}\,^{\gamma}$ and $g^{\gamma}$ distributions 
	in LO and NLO(DIS$_{\gamma}$) at fixed values of $Q^2$.  The 
	dashed--dotted curves show the hadronic NLO contribution 
	$f_{\rm{had}}^{\gamma}$ to $f^{\gamma}$.  The valence--like inputs
	at $Q^2= Q_0^2 \equiv \mu_{\rm{LO,\,NLO}}^2$ \cite{GRV98}, according to 
        eq.\ (\ref{input}), are
	shown by the lowest curves referring to $\mu^2$.  
        For comparison
	we show the steeper NLO GRV$_{\gamma}$ \cite{GRVg92} expectations
	as well.  The results have been multiplied by the number 
	indicated in brackets.}
\label{fig3}
\end{figure}
\section{Numerical Results}

\subsection{Real Photon}
In Fig.\ \ref{fig1b} we compare our LO and NLO predictions for $F_2^{\gamma}$ 
,given by the full and dashed lines, with experimental data \cite{Data99}.
Also shown are the GRV$_{\gamma}$ \cite{GRVg92} predictions which are 
steeper in the small-$x$ region (due to the steeper sea, which has been generated
from a vanishing input).
In all cases the charm contribution has been added according to a 
fixed order calculation (see, e.g., eqs. (15) and (16) in \cite{GRSvg99}).

As one can see, we achieve a good description of the data, 
particularly in the region
of small $Q^2$ where the GRV$_\gamma$ and also the SaS 1D \cite{SaS95} expectations 
fall somewhat below the data.
Furthermore, our LO and NLO results are close 
together demonstrating an excellent perturbative stability.

Figure \ref{fig3} shows the $x$--dependence of the up quark
and the gluon for three values of $Q^2$.
The solid and dashed lines are our NLO and LO results, respectively. 
The dotted lines are the NLO GRV$_{\gamma}$ distributions.

Comparing the solid lines with the dashed--dotted lines, representing 
the hadronic
contribution to the solid lines, one can see that
for $x \gtrsim 0.1$ the pointlike part dominates (especially for the
quark distribution),
whereas at very small x the hadronic component is dominant implying
a small--$x$ behaviour similar to the pion or proton. 

Also shown are our valence--like inputs which
are (vanishingly) small for $x \lesssim 10^{-2}$. 
Thus, the small--$x$ increase at higher scales 
is purely due to the QCD evolution.

\begin{figure}[bht]
\vspace*{-0.25cm}
\epsfig{figure=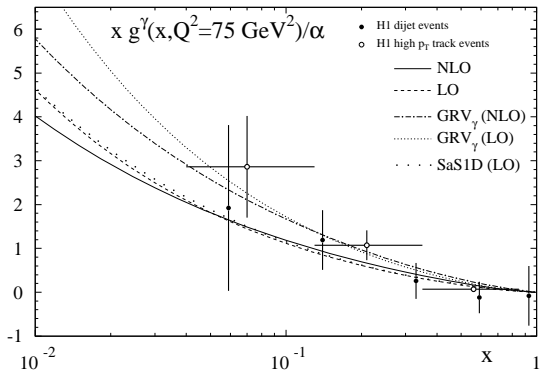,width=7.5cm}

\vspace*{-0.8cm}
\caption{\small
        Comparison of our LO and NLO predictions for
	$xg^{\gamma}$ at $Q^2\equiv \langle (p_T^{\rm{jet}})^2 \rangle
	= 75$ GeV$^2$ with HERA(H1) measurements \cite{gluon}.  The
	GRV$_{\gamma}$ and SaS expectations are taken from refs.\ 
	\cite{GRVg92} and \cite{SaS95}, respectively.}
\label{fig4}
\end{figure}
In Fig.\ \ref{fig4} we compare the $x$-dependence of our gluon distribution 
with recent H1 measurements \cite{gluon}.
The full data points are dijet events and the open points are 
high $p_T$ track events.
Our LO and NLO results are very similar to the gluon of SaS 1D, shown
by the dotted line.
The flatter small-$x$ behaviour compared to GRV$_{\gamma}$ is partly 
induced by the flatter gluon distribution in the GRV-98 proton \cite{GRV98}.
(Another reason is that the GRV$_\gamma$ gluon is enhanced by a factor 
$\kappa = 1.6 (2.0)$ in NLO (LO) at $Q_0^2$ in eq.\ (2) in \cite{GRVg92}.)
\begin{figure}[htb]
\epsfig{figure=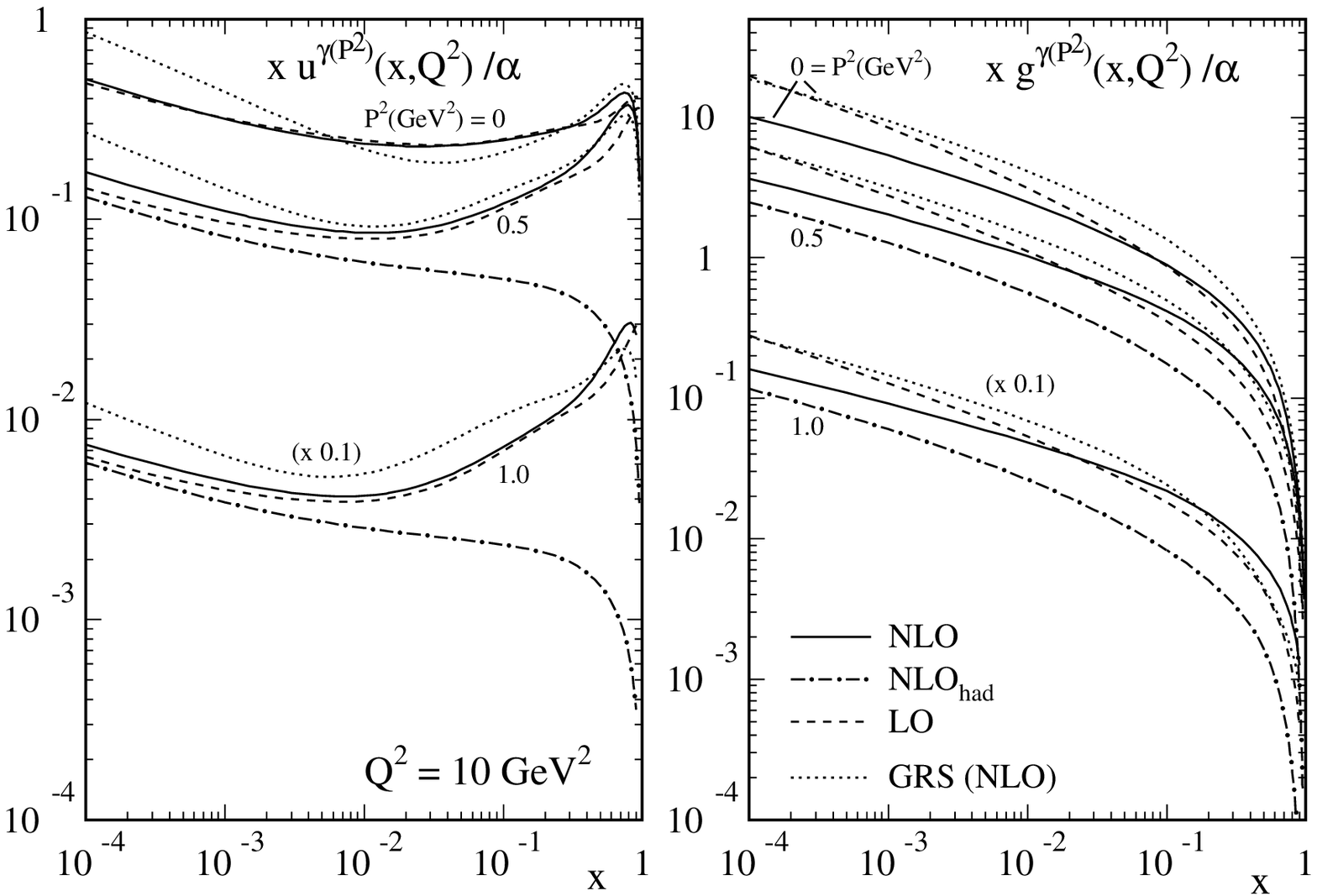,angle=0,width=7.5cm}
\epsfig{figure=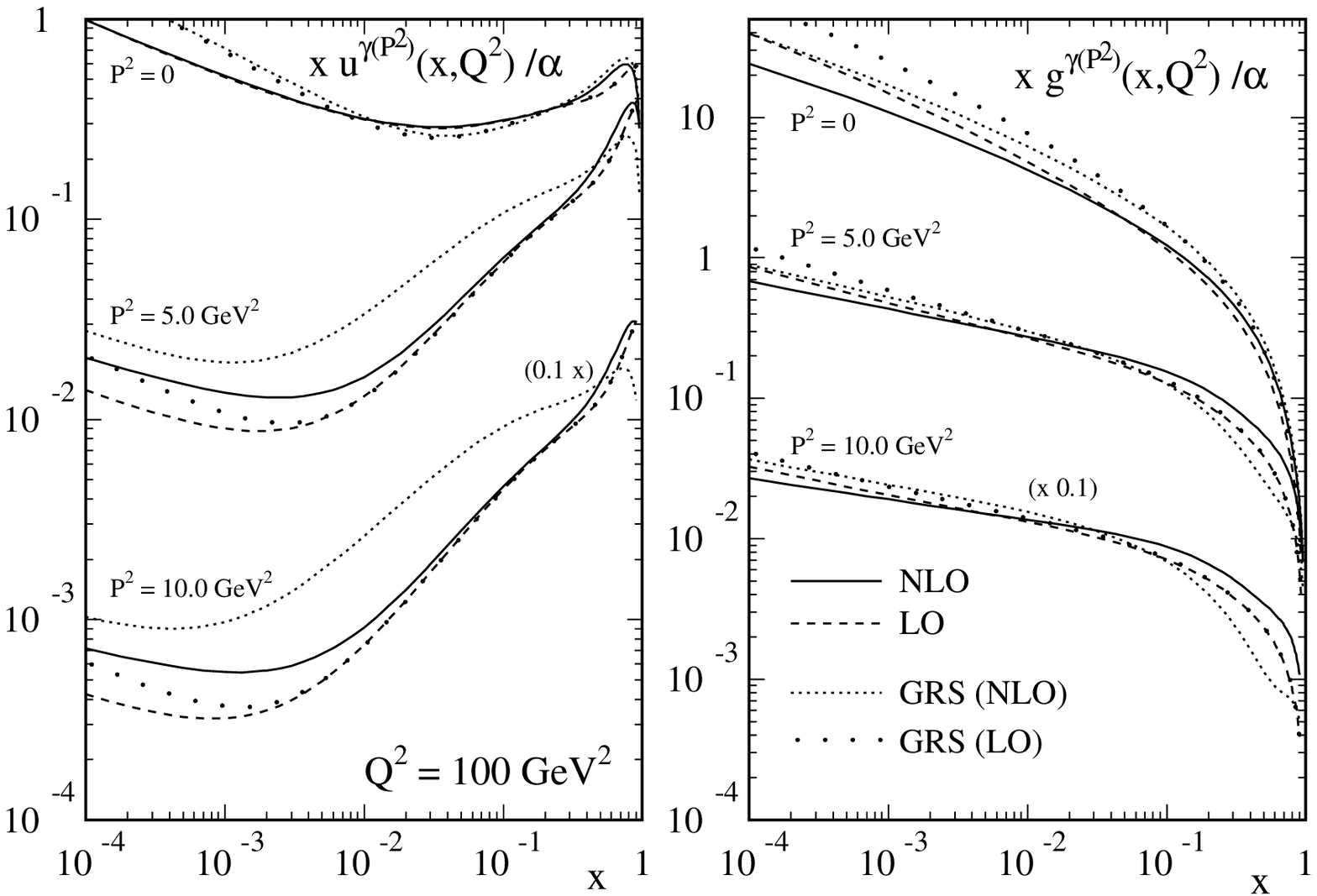,angle=0,width=7.5cm}

\vspace*{-0.8cm}
\caption{\small
        LO and NLO (DIS$_\gamma$) predictions for the up--quark and 
	gluon distributions
	of a virtual photon $\gamma(P^2)$ at $Q^2=10$ and $100$ GeV$^2$.  For 
	comparison the results for the real photon ($P^2=0$) are shown as
	well.  For $P^2=0.5$ and $1.0$ GeV$^2$ the NLO `hadronic' contribution 
        in (\ref{plhad}) is also shown
	separately.  The GRS expectations are taken from ref.\ \cite{GRS95}.
	The results have been multiplied by the numbers indicated
	in brackets.}
\label{fig9}
\end{figure}

\subsection{Virtual Photon}
Figure \ref{fig9} is the virtual photon analogue of Fig.\ \ref{fig3}.
Again the $x$-dependence of the up quark and the gluon 
is shown for two values of $Q^2$ and several photon virtualities $P^2$.
The solid and dashed curves are our NLO and LO distributions, respectively, 
whereas the dotted curves
are the NLO predictions of Gl{\"u}ck, Reya and Stratmann 
\cite{GRS95}, denoted by GRS.
For $Q^2 = 100$ GeV$^2$ we also show the LO GRS up quark distributions (wide
dotted), which are very similar to our LO results.

\begin{figure}
\epsfig{figure=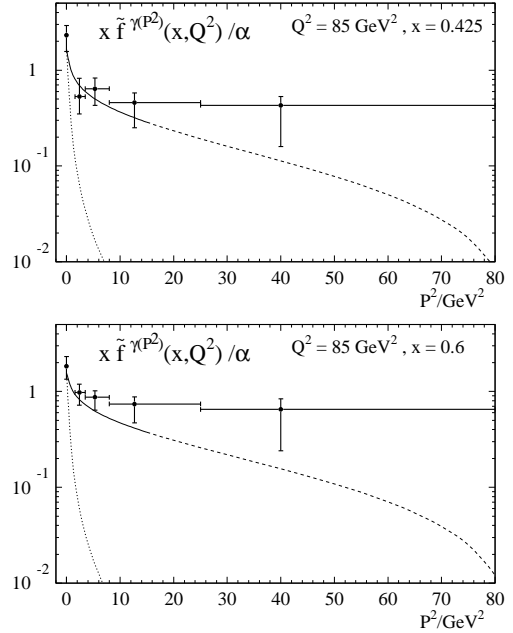,angle=0,width=7.5cm}

\vspace*{-0.8cm}
\caption{\small
        Predictions for the LO effective parton density 
	$x\tilde{f}\,^{\gamma(P^2)}(x,Q^2)$ at
	the scale $Q^2\equiv\left(p_T^{\rm{jet}}\right)^2=85$ GeV$^2$
	and at two fixed values of $x$.  The H1 data \cite{feff99} have
	been extracted from DIS ep dijet production.  The solid curves
	refer to our predictions in the theoretically legitimate region
	$P^2\ll Q^2\equiv\left( p_T^{\rm{jet}}\right)^2$, whereas the
	dashed curves extend into the kinematic region of larger $P^2$
	approaching $Q^2$ where the concept of parton distributions of
	virtual photons is not valid anymore (see text).}
\label{fig10}
\end{figure}

On can see that the pointlike part becomes increasingly important with increasing
$P^2$ due to the dipole suppression of the hadronic component
[$\eta(P^2 ({\rm GeV}^2) = 0.5,1.0,5.0,10.0) = 0.3, 0.14, 0.01, 0.003$].
At $P^2 = 5 , 10$ GeV$^2$ the hadronic part is nearly 
vanishing even at small $x$.
Furthermore, note the perturbative stability of the parton distributions 
guaranteeing the required stability of the physical structure functions.

Finally, in Fig.\ \ref{fig10} we compare our LO parton distributions with an 
effective parton density
\begin{eqnarray}
\tilde{f}\,^{\gamma(P^2)}(x,Q^2) &=& \sum_{q=u,d,s}\left(q^{\gamma(P^2)}
   + \bar{q}\,^{\gamma(P^2)}\right)\nonumber\\
&+&\, \frac{9}{4}\,g^{\gamma(P^2)}
\end{eqnarray}
which
has been extracted from H1 dijet data \cite{feff99}.
$Q \equiv p_T^{\rm{jet}}$ and we show results for two values of $x$ 
in dependence of the photon virtuality.
The transition from the solid to the dashed line indicates that for
$P^2$ approaching $Q^2$ the concept of renormalization group resummed 
parton distributions of
virtual photons is not expected to hold any more, because the resummed logarithms
$\log Q^2/P^2$ arising in the partonic subprocesses are not much 
larger than the non-logarithmic terms.

In the relevant kinematic region $P^2 << Q^2$
the agreement with the data is reasonably good.
This is in contrast to a simple dipole suppression of the real photon
$\eta(P^2) \tilde{f}\,^{\gamma(P^2=0)}(x,Q^2)$, as illustrated by the 
dotted curves in Fig.\ \ref{fig10}, which fails to describe the data.

\vspace*{0.5cm}
\noindent
Supported in part by the {\it Gra\-du\-ierten\-kolleg 'Erzeugung und Zer\-f\"alle 
von Ele\-mentar\-teil\-chen'} of the {\it Deutsche Forschungs\-gemein\-schaft} 
at the {\it Uni\-versit\"at Dortmund} and by the {\it Bundes\-ministerium f\"ur
	Bildung, Wissenschaft, Forschung und Technologie}, Bonn.

\end{document}